\documentclass[11pt,twoside]{article}
\usepackage{amsmath,epsfig,graphicx,amssymb}

\newcommand{\be}{\begin{equation}}
\newcommand{\ee}{\end{equation}}

\newcommand{\scal}[2]{\langle #1\left|#2\rangle\right.}
\newcommand{\vlmd}{\ensuremath{\boldsymbol{\lambda}}}

\newcommand{\vc}[1]{\ensuremath{\boldsymbol{#1}}}
\newcommand{\nbl}[1]{\nabla_{\!\! #1}}

\def\cU{\ensuremath{U(\mathcal{C})}}
\def\A{\ensuremath{\boldsymbol{A}}}
\def\diag{\ensuremath{\mathop{\rm{diag}}}}

\def\bsm{\boldsymbol}

\begin{document}

\title{\bf Application of Geometric Phase in Quantum Computations}

\author{A.E. Shalyt-Margolin, V.I. Strazhev and A.Ya. Tregubovich\\
National Centre of High Energy Physics,\\ 220040 Bogdanovicha str.
153, Minsk, Belarus\\ E-mail: alexm@hep.by, a.tregub@open.by}
\date{}

\maketitle
{\bf \Large Computer Science and Quantum Computing,
p.p.125-149, Nova Science Publishers 2007} \\
\\{\Large \sl  This
article is dedicated  to memory of our dear friend, colleague and
co-author Dr. Artur Tregubovich, (1961--2007)}
\begin{abstract}
Geometric phase that manifests itself in number of optic and nuclear
experiments is shown to be a useful tool for realization of quantum
computations in so called  holonomic quantum computer model (HQCM).
This model is considered as an externally driven quantum system with
adiabatic evolution law and finite number of the energy levels. The
corresponding evolution operators represent quantum
 gates of HQCM. The explicit  expression for the gates
 is derived both for one-qubit and for multi-qubit quantum gates as Abelian and non-Abelian
 geometric phases provided the energy levels to be time-independent or in other words for
 rotational adiabatic evolution of the system. Application of non-adiabatic geometric-like phases
 in quantum computations is also discussed for a Caldeira-Legett-type model (one-qubit gates)
 and for the spin 3/2 quadrupole NMR model (two-qubit gates). Generic quantum gates for these two
 models are derived. The possibility of construction of the universal quantum gates in both cases
 is shown.
\end{abstract}


\thispagestyle{myheadings}

\section{Introduction}

The conceptions of quantum computer (QC) and quantum computation developed in
80-th \cite{Deutsch1}, \cite{Feynman} were found to be fruitful both for
computer science and mathematics as well as for physics \cite{Steane}. Although
a device being able to perform quantum computations is now far away from practical
realization, there is a great number of theoretical proposals of such a construct
(see e.g. \cite{Lloyd1}--\cite{Gershenfeld}).
 Intensive investigations on quantum information theory (see e.g.  \cite{cabello}, \cite{kilin1}
 for a reference source on this subject) refreshed some interest on Berry phase
 \cite{Berry}. The idea of using unitary transformations produced by
 Berry phase  as quantum computations is proposed in
\cite{Zanardi1}, \cite{Zanardi2} and first realized in \cite{Pachos1}, \cite{pachos3} in
 a concrete model of holonomic quantum computer where the degenerate states of laser beams in non-linear
 Kerr cell are interpreted as qubits.  For other references where Abelian Berry phase is considered
in the context of quantum computer see e.g. \cite{ekert} - \cite{averin}. If the corresponding energy
level is degenerate non-Abelian phase takes place \cite{Wilczek-Zee} that is actually a matrix
mixing the states with the same energy. For further references on quantum computation based on
non-Abelian geometric phase see e.g. \cite{pachos3}--\cite{yi}.

  On the other hand non-adiabatic analogue of Berry phase can exist and be
measured  if transitions in a given statistical ensemble do not lead to loose of
coherence \cite{appelt1}. For loose of coherence in quantum computations related to geometric phase
see \cite{nazir}--\cite{yi2}.  Thus it is also possible to use the corresponding
unitary operators to realize quantum gates. This fact has been noticed
in \cite{keiji1}, \cite{keiji2}. After that a lot of papers was published where the
non-adiabatic phase is applied
to realize the basic gates in different models of QC such as different NMR schemes
\cite{keiji3}--\cite{das}, ion traps \cite{solinas1},
\cite{solinas2}, quantum dots \cite{li}, \cite{scala}, and superconducting nanocirquits
\cite{wang2}.

To analyze a concrete scheme for quantum computation based on geometric phase it is desirable
to be aware of analytical expression for the evolution operator of the system at least at the
moment when the measurement is performed. This article is concentrated on the computational  aspect
of geometric phase for the models which are relevant to QC. It should be emphasized that the form
of the expression for the phase and the possibility of the derivation of such a formula itself
thoroughly depend on the group-theoretic structure of the corresponding Hamiltonian. Therefore
a  method of the geometric phase calculation which would be more or less universal at least in the
adiabatic case can appear to be useful. The material is divided in two parts. In section \ref{adiab} the
adiabatic geometric phase is considered. In subsection \ref{ab_adiab} we analyze the difficulties
appearing in calculation of the Abelian adiabatic geometric phase (Berry's phase) and propose a
method of its explicit derivation for the case of the symmetric
time-dependent Hamiltonian with constant non-degenerate energy levels. The symmetry of the Hamiltonian is supposed
to reduce the Hamiltonian to that of a system with finite number of energy levels. In subsection
\ref{nonab_adiab} this method is generalized for the case when degeneration is present. In section
\ref{nonadiab} we consider non-adiabatic phase which can only conditionally be called
"geometric" for its dependence on concrete details of the dynamics. For this reason it is not
possible to work out more or less general approach to the calculation of the non-adiabatic phase.
Therefore two concrete cases are considered. In subsection \ref{ab_nonadiab} application of the
Abelian non-adiabatic phase to one-qubit computation in a Caldeira-Legett-type model is cosidered.
 In subsection
\ref{nonab_nonadiab} we present an example of both non-Abelian and non-adiabatic phase computation
in spin-3/2 quadrupole NMR resonance model.

\section{Adiabatic Geometric Phase}\label{adiab}
\subsection{Abelian Berry's Phase}\label{ab_adiab}

Here we consider a possible method of the adiabatic phase computation that seems to be effective
in a broad range of practically relevant cases. Berry phase is a consequence of the adiabatic
(or Born--Fock)  theorem \cite{Messiah}
which states that a parametric quantum system depending on a set of slowly (adiabatically) evolving
parameters $R_i(t),\quad i=1,\ldots N$ behaves in a quasi-stationary manner
\begin{equation}\label{AdiabSchroed}
\hat{H}(\vc{R} ) |n(\vc{R} )> = E_n(\vc{R} ) |n(\vc{R} )>,
\qquad   \vc{R} = (R_1,\ldots ,R_N)
\end{equation}
where $\hat{H}(\vc{R})$ is the corresponding Hamiltonian and no energy level degeneration is assumed.
The adiabaticity condition means that the
frequencies $\omega_n(\vc{R}) = E_n(\vc{R})/\hbar$ are much greater than the characteristic Fourier
frequencies of $R_i(t)$. Thus the eigenvectors $|n(\vc{R})>$ evolve like
\begin{equation}\label{Sdef}
|n(\vc{R})> = \hat{S}(\vc{R}) |n_0> , \qquad |n_0> = |n(\vc{R}(0))>, \quad \hat{S}\hat{S}^{\dagger} = 1
\end{equation}
with unitary rotation $\hat{S}$ describing the natural variation of $|n(\vc{R})>$ due to that
 of $\vc{R}(t)$. It corresponds to the following evolution law of the Hamiltonian
\begin{equation}\label{SRHevol}
  \hat{H}(t) = \hat{S}(\vc{R} )\,\hat{H}_0(t)\, \hat{S}^{\dagger}(\vc{R})
\end{equation}
where $H_0(t)$ is diagonal in the basis $\{|n_0> \}$.
  What is the solution of the non-stationary Schr\"{o}dinger equation
\begin{equation}\label{genSchroed}
  i\hbar\,\frac{\partial}{\partial t}|\psi (t)> = \hat{H}(t) |\psi (t)>
\end{equation}
for this case? A natural hypothesis would be that the evolution operator for $|\psi >$
$$
     |\psi (t)> = \hat{U}(t)|\psi (0)>
$$
has the form
$$
   \hat{U}(t) = \hat{S}(\vc{R})\hat{\Phi}(t)
$$
where $\hat{S}$ is defined by (\ref{Sdef}) and $\hat{\Phi}$ simply produces the dynamic phase
\begin{equation}\label{defPhi}
    \hat{\Phi}(t) |n(\vc{R} (t)> = \exp\left(-i/\hbar\,\int\limits_0^t E_n(\tau )\,
                                d\tau\right)\, |n(\vc{R} (t)>.
\end{equation}
It is based on the analogue with the stationary case where evolution is simply represented by the
dynamical phase factor $\exp (-i/\hbar\, E_n\, t)$. Berry first observed \cite{Berry} that the
hypothesis is wrong. To see this it is sufficient to represent $\hat{U}$ in the form
$\hat{U}(t) = \hat{S}(\vc{R} )\hat{V}(t)$ and substitute it into the Schr\"{o}dinger equation
\begin{equation}\label{USchroed}
  i\hbar\, \frac{\partial}{\partial t}\hat{U}(t) = \hat{H}(t) \hat{U}(t).
\end{equation}
It gives
\begin{equation}\label{Vdot}
  i\hbar\,( \hat{\dot{V}}\hat{V}^{\dagger} + \hat{S}^{\dagger}\nabla_{\!\!\vc{R}}\hat{S}\, \dot{\vc{R}} )
                  = \hat{H}_0(t).
\end{equation}
Now one can see that $\hat{V}$ cannot be simply $\hat{\Phi}$ because it has to cancel the second
term in the right hand side of (\ref{Vdot}) besides of $H_0$. It follows from (\ref{Vdot})that
$$
    \hat{V}(t) = \hat{\Gamma}(t)\hat{\Phi}(t)
$$
where $\hat{\Phi}(t)$ is determined by (\ref{defPhi}) and the following equation is valid for
$\hat{\Gamma}(t)$:
\begin{equation}\label{defGamma}
  \hat{\dot{\Gamma}}\hat{\Gamma}^{\dagger}\, |n_0> = - (\hat{S}^{\dagger}\nbl{\vc{R}}\hat{S})
                                                       \dot{\vc{R}}\,\, |n_0>.
\end{equation}
It results in the evolution law for the state vector corresponding to the $n$-th energy level
 \begin{equation}\label{defAbPhase}
 |\psi_n (t)> = e^{-i/\hbar\, \Phi_n (t)}\, e^{i\gamma_n (t)}\,\, |n(\vc{R} (t))>
 \end{equation}
where $\Phi_n(t)$ is the phase factor in the right-hand side of (\ref{defPhi}) and $\gamma_n(t)$
is given by
 \begin{equation}\label{defgamma}
 \gamma_n (t) = \int\limits_0^t\!\! \vc{A}_n(\vc{R}(\tau ))\dot{\vc{R}}(\tau )\,\, d\tau, \quad
                    \vc{A}(\vc{R}) = i <n(\vc{R} )|\,\nbl{\vc{R}} n(\vc{R} )>.
 \end{equation}
Phase $\gamma_n (t)$ becomes purely geometric while $\vc{R}(t)$ evolves cyclically:
$\vc{R}(T) = \vc{R}(0)$
\begin{equation}\label{defPBer}
  \gamma_n (T) \equiv \gamma_n(\mathcal{C}) = \oint\limits_{\mathcal{C}}\!\! \vc{A}_n (\vc{R})\, d\vc{R}.
\end{equation}
Here the integration contour $\mathcal{C}$ is a closed curve in the parameter space described by
$\vc{R}(t)$ as a radius-vector. It is easily seen from (\ref{defPBer}) that $\gamma_n(\mathcal{C})$
does not depend on the concrete details of the system's dynamic if the adiabatic condition is held.

  In this article we are interested in computing of $\gamma_n(\mathcal{C})$ in the most general case.
  The problem of derivation of $\gamma_n(\mathcal{C})$
was solved  in various particular cases in large number of articles some years ago. First we would like
to note that the straightforward formula
\begin{equation}\label{Bn}
  \vc{B}_n = \nbl{\vc{R}}\times \vc{A}_n = \sum\limits_{m\neq n}\,\frac{(\nbl{\vc{R}}\hat{H}(\vc{R}))_{mn}\times
                                         (\nbl{\vc{R}}\hat{H}(\vc{R}))_{nm}}{(E_n(\vc{R}) - (E_m(\vc{R}))^2}
\end{equation}
derived by Berry \cite{Berry} by making use of the identity
$$
   <m|\nabla n> = \frac{<m|\nabla \hat{H} |n>}{(E_n - E_m)}, \quad m \neq n
$$
has not (despite of it's beauty) much practical use because to apply it one should establish the
analytical dependence of all $E_n$ on $\vc{R}$ that is not a realistic task excluding some special
cases. To see this one should attempt to apply formula (\ref{Bn}) to the case of a 3-level system
substituting a generic solution $E_n(\vc{R})$ of the corresponding cubic equation therein.

It was first noticed in \cite{AnandStod} that symmetries of the Hamiltonian $\hat{H}(\vc{R})$
play an important role in computing of $\gamma_n$. Indeed if one represents the result of the
periodic motion $\hat{H}(0) = \hat{H}(T)$ as
\begin{equation}\label{UT}
  |\psi (T)> = \hat{U}(T) |\psi (0)>
\end{equation}
where as it follows from (\ref{AdiabSchroed}) $\hat{U}(T)$
must commute with $\hat{H}(0)$ so it can be represented as an exponent containing a linear
combination of operators $\hat{X}_k$ which must commute with $\hat{H}(0)$ as well. Thus the operators $\hat{X}_k$
are integrals of motion and describe certain symmetries of the given system. Therefore in what follows
we restrict ourselves with such systems whose Hamiltonian is an element of a finite Lie algebra. This
assumption immediately gives the group-theoretic structure of $\hat{U}(T)$:
\begin{equation}\label{UTLie}
  \hat{U}(T) = \exp\left(i\,\sum\limits_i\, a_i\, H_i\right)
\end{equation}
where $H_i$ are all linearly independent elements of the Cartan subalgebra and $a_i$ are some
coefficients. Thus the problem reduces to computing of the coefficients $a_i$. In the simplest
case of Lie algebras consisting of three elements this problem can be easily solved \cite{Tregub1},
\cite{sriram}, \cite{chiao} for physically relevant cases of Heisenberg-Weyl algebra, $su(2)$ and
$su(1.1)$. In each of them the evolution operator has the form
\begin{equation}\label{UgA3}
  \hat{U}(t) = \exp\left( \zeta (t) \hat{X}_+ - \zeta^*(t) \hat{X}_-\right)\,
               \exp\left(i\,\phi (t)\, \hat{X}_3\right)
\end{equation}
provided the initial Hamiltonian is proportional to $\hat{X}_3$ where $\hat{X}_{\pm}$ and
$\hat{X}_3$ are the corresponding generators of the algebras above. Their expressions for each
concrete case are given in table \ref{eXpr}.
\begin{table}[tb]
\caption{Expressions for the operators $\hat{X}_{\pm},\,
\hat{X}_3$.}\label{eXpr}
\begin{center}
\begin{tabular}{|c|c|c|c|l|}
\hline
Algebra & \raisebox{-2pt}{$\hat{X}_+$} & \raisebox{-2pt}{$\hat{X}_-$}
                              & \raisebox{-2pt}{$\hat{X}_3$} & Commutators \\
\hline
$su(2)$ & $\hat{J}_+ $  & $\hat{J}_-$ & $\hat{J}_3$  &  \begin{tabular}{l}
                                          \raisebox{-10pt}{$[\hat{J}_3, \hat{J}_{\pm}] =
                                                        \pm\hat{J}_{\pm}$}\\
                                                             \\
                                          \raisebox{7pt}{$[\hat{J}_+, \hat{J}_-] = 2\hat{J}_3$}
                                                        \end{tabular}\\
\hline
$su(1.1)$ & $\hat{K}_+ $  & $\hat{K}_-$ & $\hat{K}_3$ & \begin{tabular}{l}
                                           \raisebox{-10pt}{$[\hat{K}_3, \hat{K}_{\pm}] =
                                                        \pm\hat{K}_{\pm}$}\\
                                                               \\
                                           \raisebox{7pt}{$[\hat{K}_+, \hat{K}_-] = -2\hat{K}_3$}
                                                        \end{tabular}\\
\hline
$H-W$ & $\hat{a}^+ $  & $\hat{a}$ & $\hat{1}$  &  \begin{tabular}{l}
                                            \raisebox{-10pt}{$[\hat{1}, \hat{a}^+] =
                                                        [\hat{1}, \hat{a}] =0$}\\
                                                         \\
                                            \raisebox{5pt}{$[\hat{a}, \hat{a}^+] = \hat{1}$}\\
                                            \end{tabular}\\
\hline
\end{tabular}
\end{center}
\end{table}
The Hamiltonian for $su(2)$ case describes an arbitrary spin in the magnetic field so all $J$'s are
the angular momentum operators: $\hat{J}_{\pm} = 1/2(\hat{J}_1 \pm \hat{J}_2)$.
 $su(1.1)$ case corresponds to the evolution of squeezed states
\cite{Stoler} of light in non-linear optics. Here $\hat{K}_+ = \hat{a}^{+^{2}}/2$,
$\hat{K}_- = \hat{a}^2/2$ and $\hat{K}_3 = \hat{a}^+\hat{a} + 1/2$ where $\hat{a}$, $\hat{a}^+$ are
usual bosonic annihilation and creation operators. The last case represents a harmonic oscillator
interacting with the time-dependent electric field. The simple commutation relations in these three
algebras admit direct computation of Berry's phase \cite{Tregub1}, \cite{sriram}, \cite{chiao}.
\begin{equation}\label{3elgamma}
  \gamma_m = m\,\oint\limits_{\mathcal{C}}\omega (\xi ) = \int\limits_S d\wedge \omega (\xi )
\end{equation}
where $m$ is an eigenvalue of the corresponding $\hat{X}_3$, $S$ is the surface in the parameter
space bounded by the closed curve $C$ and the expressions for $\omega (\xi )$ and its external
derivative $d\wedge
\omega (\xi )$ are given for each case in table \ref{omega}.
\begin{table}
\caption{Expressions for the forms $\omega$ and $d\wedge\omega$.}
\label{omega}
\begin{center}
\begin{tabular}{|l|c|c|c|}
\hline
  Algebra & $\omega (\xi )$ & $d\wedge\omega (\xi )$ & Relation to $\zeta$  \\
\hline
  \raisebox{-10pt}{$su(2)$}   & \raisebox{-10pt}{$\frac{{\displaystyle\xi d \xi^* - \xi^* d \xi}}
                              {{\displaystyle 1 + |\xi |^{2^{\vphantom{1}}}}}$} &
              \raisebox{-10pt}{$\frac{{\displaystyle 2\, d\xi \wedge d\xi^*}}
              {{\displaystyle (1 + |\xi |^{2^{\vphantom{1}}})^{2^{\vphantom{1}}}}}$} &
              \begin{tabular}{c}
              \raisebox{-10pt}{$|\xi | = \tan (|\zeta |),$} \\
              \raisebox{-10pt}{$\arg\xi = \arg\zeta$}
                               \end{tabular} \\
            &         &      &  \\
\hline
  \raisebox{-10pt}{$su(1.1)$} & \raisebox{-10pt}{$\frac{{\displaystyle\xi d \xi^* - \xi^* d \xi}}
              {{\displaystyle 1 - |\xi |^{2^{\vphantom{1}}}}}$} &
              \raisebox{-10pt}{$\frac{{\displaystyle 2\, d\xi \wedge d\xi^*}}
              {{\displaystyle (1 - |\xi |^{2^{\vphantom{1}}})^{2^{\vphantom{1}}}}}$} &
              \begin{tabular}{c}
              \raisebox{-10pt}{$|\xi | = \tanh (|\zeta |),$} \\
              \raisebox{-10pt}{$\arg\xi = \arg\zeta$}
                               \end{tabular} \\
            &         &     &    \\
\hline
  \raisebox{-10pt}{H-W}   & \raisebox{-10pt}{$\xi d \xi^* - \xi^* d \xi$} &
                                                   \raisebox{-10pt}{$2\, d\xi \wedge d\xi^*$} &
         \raisebox{-10pt}{$\xi = \zeta $} \\
        &             &     &     \\
\hline
\end{tabular}
\end{center}
\end{table}
The geometric sense of the derived phase factor is the integral curvature over the surface
bounded by the contour
$\mathcal{C}$ on the manifold the evolution operator belongs to. This manifold can be generally
expressed in the form $G/H$ where $G$ is the group manifold and $H$ is that of the stationary
subgroup, i.e. the group whose Lie algebra consists of all operators commuting with $H(0)$ (in
these three cases it is always $U(1)$).
It is sphere in the case
of $su(2)$, two-sheet hyperboloid  in the case of $su(1.1)$ and plane in the case of Heisenberg-Weyl
group. To complete the computation one has to establish correspondence between the complex
parameter $\xi$ and physical parameters of the Hamiltonian. Let us do that for $SU(2)$. It is worth
to notice that the result is completely determined by the geometric properties of the group and does
not depend on the concrete representation. For this reason one can chose the fundamental representation
of $SU(2)$ to simplify the derivation. Thus we take $H(0) = \omega_B\, \sigma_3$ which
corresponds to the initial eigenvectors $|\pm> = (1 (0), 0 (1) )^T$ ($T$ denotes transposition).
The evolution operator generally parametrized by the spherical as
\begin{equation}\label{Uteta}
  \begin{pmatrix}
    \cos\!\frac{\,\,\vartheta}{\,2\vphantom{)^A}} &
                                     -\sin\!\frac{\,\,\vartheta}{\,2\vphantom{)^A}}\, e^{-i\varphi } \\
    \vphantom{s}   &  \vphantom{s} \\
    \sin\!\frac{\,\,\vartheta}{\,2\vphantom{)^A}}\, e^{i\varphi } &
                                                       \cos\!\frac{\,\,\vartheta}{\,2\vphantom{)^A}} \\
  \end{pmatrix}
\end{equation}
rotates $H(0)$ into $H(t) = \omega_B\, \vc{n}\vc{\sigma}$ where
$\vc{n} = (\cos \varphi\sin\vartheta /2, \,\,\sin\varphi\sin\vartheta /2, \,\,\cos\vartheta /2 )$ determines
the direction of the magnetic field.
On the other hand the direct computation of $\exp (\zeta \hat{J}_+ - \zeta^*\hat{J}_-)$ gives
for this representation $|\zeta | = \vartheta /2, \,\arg\zeta = \varphi + \pi$. It leads to the well
known expressions for the fictitious "strength field"
\begin{equation}\label{Bpm}
  \vc{B}_{\pm} = \mp\,\frac{\raisebox{-4pt}{1}}{2}\, \frac{\raisebox{-4pt}{\vc{R}}}{R^3}
\end{equation}
where $\vc{R}$ denotes the true magnetic field vector in order not to confuse it with the
fictitious one which determines the resulting Berry's phase. It should be noted that the
correspondence $\vc{R} \rightarrow \xi$ realizes the stereographic projection of the sphere with
the coordinates $\vartheta, \varphi$ on the plane that points are labeled by $\xi$.
 The other two cases of $SU(1.1)$
and Heisenberg-Weyl groups can be considered in a similar manner.

Re-derivation of these simplest results has the intention to extract a universal idea of
computing the geometric phase in more or less general case. For the sake of certainty let
us assume the symmetry algebra of the Hamiltonian to be semisimple. It means that the generic
evolution operator can be represented in the form
\begin{equation}\label{UProdAlf}
  \hat{U}(t) = \prod\limits_{\alpha\in\,\Delta_+}\, \hat{U}_{\alpha}(t)
\end{equation}
where $\Delta_+$ denotes the set of the positive roots $\alpha$ and each $U_{\alpha}$ is
analogous to (\ref{UgA3}) (see also table \ref{eXpr} for $su(2)$ and $su(1.1)$ cases):
\begin{equation}\label{UAlf}
  \hat{U}_{\alpha}(t) = \exp\left( \zeta_{\alpha}(t)\, \hat{E}_{\alpha} \, - \,
                                   \zeta^*_{\alpha}(t)\, \hat{E}_{-\alpha}\right)
\end{equation}
where the standard notations for the Cartan basis \cite{Goto}
\begin{equation}\label{CrtBasis}
  [H_{\beta}, \hat{E}_{\pm\alpha}] = \pm\alpha (H_{\beta})\,\hat{E}_{\pm\alpha}\quad
  [\hat{E}_{\alpha}, \hat{E}_{-\alpha}] = H_{\alpha},\quad
  [\hat{E}_{\alpha}, \hat{E}_{\beta}] = N_{\alpha\beta}\,\hat{E}_{\alpha + \beta}
\end{equation}
are used. The pairs of generators $\hat{E}_{\pm\alpha}$ are analogous for $\hat{J}_{\pm}$ in
$su(2)$ and $H_{\alpha}$'s are that of $\hat{J}_3$. Here $\alpha (H_{\beta})$ and $N_{\alpha\beta}$
are constants that can be chosen rational and integer correspondingly. Taking account of the
consideration above leads to some more detailed form for one-cycle evolution operator
$\hat{U}(T)$ (\ref{UTLie})
\begin{equation}\label{GamAlf}
  \hat{U}(\mathcal{C}) = \exp\left( i\,\sum\limits_{\alpha\in\Delta_+}\, a^{\alpha}(\mathcal{C})
                                     H_{\alpha}\right).
\end{equation}
Each pair $(\hat{E}_{\alpha}, \hat{E}_{-\alpha})$ makes besides of the trivial group-theoretic
contribution $H_{\alpha}$ which produces the corresponding quantum number also a non-trivial one
reflecting adiabatic dynamics of the system
\begin{equation}\label{ThetAlfa}
  a^{\alpha}(\mathcal{C}) = \oint\limits_{\mathcal{C}}\,\theta^{\alpha}(\vc{\zeta})
\end{equation}
where $\theta_{\alpha}$ generally depends on all $\zeta_{\alpha}(t)$. Thus to solve the problem one
has to find this dependence making use of commutation relations (\ref{CrtBasis}) and then establish the
connection between the parameters $\zeta_{\alpha}(t)$ and the natural set of parameters $\vc{R}$ of
the Hamiltonian. Unfortunately the hope to obtain a solution of even one of these two tasks that
would be a non-trivial generalization of the  above examples is not realistic. Neither the first
part of the problem nor the second one could be solved in a way resulting in physically relevant
explicit formulas having practical use. First the 1-forms $\theta_{\alpha}$ fulfill Maurer-Cartan
equations that express the quantity $\hat{U}^{\dagger}\, d\hat{U}$
in terms of the 1-forms $\omega_{\alpha}$ and $\theta_i$
\begin{equation}\label{UdUomega}
  \hat{U}^{\dagger}(\vc{\zeta})\, d\hat{U}(\vc{\zeta}) = i(\omega_{\alpha}(\vc{\zeta})\,
          \hat{E}_{\alpha} + \theta_i(\vc{\zeta})\, H_i)
\end{equation}
where the index $i$ labels all linearly
independent generators of the Cartan subalgebra (not all $H_{\alpha}$ are so). For commutation
relations (\ref{CrtBasis}) these equations take the form
\begin{eqnarray}
   d\wedge\omega^{\alpha} & = & C^{\alpha}_{\beta k}\,\,\omega^{\beta}\wedge\theta^k  +
                           1/2\,\, C^{\alpha}_{\beta\lambda}\,\,\omega^{\beta}\wedge\omega^{\lambda}
                           \label{MrCrt11}     \\
   d\wedge\theta^i        & = & 1/2\,\, C^i_{\beta\lambda}\,\,\omega^{\beta}\wedge\omega^{\lambda}
                           \label{MrCrt12}
\end{eqnarray}
Equations (\ref{MrCrt11}), (\ref{MrCrt12}) describe the parallel transport on the coset manifold
$G/H$. The possibility to solve them depends on the manifold's symmetry and of course is entirely
determined by the structure constants $C^{\cdot}_{\cdot\,\cdot}$ that are built from the root vectors
$\alpha (H_{\beta})$ and the constants $N_{\alpha\beta}$. The general solution of this system can
be constructed for very high symmetry of symmetric spaces \cite{Helgasson} where the whole algebra
can be split in two subsets $X$ and $Y$ such that
$$
   [Y, Y] \subset Y, \quad [Y, X] \subset X, \quad [X, X] \subset Y.
$$
It is seen from (\ref{CrtBasis}) that the last condition is generally speaking not valid for our
case because not all $N_{\alpha\beta}$ are zeroes. Its geometric sense is that the considered
coset spaces $G/H$ are of more general symmetry type than symmetric spaces. Thus for $G = SU(n)$
the space
\begin{equation}\label{SUnU1n}
SU(n)/\underbrace{U(1)\times U(1)\ldots\times U(1)}_{n-1\,\,\, \rm{times}}
\end{equation}
belongs to the more general class of K\"{a}hlerian spaces. The general solution of (\ref{MrCrt11}),
(\ref{MrCrt12}) for the types of spaces we are interested in is not obtained so far. Therefore
the practical use of these equations is not high. Moreover the solution of the second part of the
problem discussed is not possible for the same reason.

A simple and effective
method of practical computation of geometric phase where it is not necessary to find the forms
$\omega^{\alpha}$, $\theta^i$ is proposed in \cite{TrAd}. For this purpose we have to make some assumptions. First we regard
the Hamiltonian to belong to a finite irreducible representation of a semisimple Lie algebra
therefore $\hat{H}_0$ in (\ref{SRHevol}) can always be represented as a finite matrix $\hat{H}_0 = \vc{R}_i(t)\,H_i$ where
the set $\{H_i\}$ is a basis of the Cartan subalgebra and $\vc{R}_i(t)$ are parameters. Then we
suppose the energy levels $E_m$ to be constants. It corresponds to a rotation-type evolution
(\ref{SRHevol}) where $\hat{H}_0$ does not depend on $t$. Such a situation takes place practically
in all experiments on the geometric phase measurement. This makes it possible to
regard $E_m$ as additional secondary parameters to be found just once (may be numerically). The
third assumption is that the spectrum remains always non-degenerate i.e., no crossing of
energy levels occurs.
As the spectrum of the Hamiltonian is finite, the state vector
$\left.\left|\varphi_m\right>\right.$ is a unit vector $\bsm{m}$ in
$\boldsymbol{\mathrm{C}}^n$, so $A_m$ is
\begin{equation}\label{abAn1}
A_m = \frac{i}{2}\, (\bsm{m}^*d\bsm{m} - \bsm{m}d\bsm{m}^*).
\end{equation}
As the evolution is adiabatic, the spectrum of $H(t)$ remains always non-degenerate
if it was so at the initial time. Then there is always a nonzero main minor of
$H - E_m$ which we assume to consist always of the first $n-1$ lines and columns of
$H - E_m$. Denoting the matrix consisting of the first $n-1$ lines and columns
of $H$ by $H_{\perp}$ we come to the condition
\begin{equation}\label{condition}
\det (H_{\perp} - E_m) \neq 0
\end{equation}
Making use of this condition one can represent $\bsm{n}$ in the uniform
coordinates
$$ \bsm{m} = \frac{(\bsm{\xi}_m, 1)}{\sqrt{1+|\bsm{\xi}_m|^2}}  $$
and express $\bsm{\xi_m}$ in terms of $H_{ij}$ for $1\le i,j\le n-1$ and $E_m$:
\begin{equation}\label{xin}
\bsm{\xi}_m = (H_{\perp} - E_m)^{-1}\, \bsm{h}, \qquad h_i = -H_{in},
\end{equation}
 where $\bsm{h}$ is a vector in $\boldsymbol{\mathrm{C}}^{n-1}$ but not in
 $\boldsymbol{\mathrm{C}}^n$. Thus we have for $A_m$
 \begin{equation}\label{abAn2}
A_m = \frac{i}{2}\, \frac{(\bsm{\xi_m}^*d\bsm{\xi_m} - \bsm{\xi_m}d\bsm{\xi_m}^*).}
 {1+|\bsm{\xi}_m|^2} ,
 \end{equation}
 where $\bsm{\xi}_m$ is completely determined by (\ref{xin}). Note that the result
 obtained is purely geometrical because it can be expressed of the K{\"a}hlerian
 potential
 $$  F = \log (1 + |\bsm{z}|^2) $$
 where $\bsm{z}$ is a vector in $\boldsymbol{\mathrm{C}}^{N^2}$ consisting of
$n(n-1)/2$ independent components of all $\bsm{\xi}_m$.
 The function $F(\vc{z}, \vc{z}^*)$ determines all the geometrical properties of the state space
 (\ref{SUnU1n}). Particularly its metric tensor is
 $$
     g_{ij} =\frac{\partial^2F(\vc{z}, \vc{z}^*)}{\partial z_i\partial z_j^*}.
$$
It should be emphasized that the simplification of the problem reached here is based on the fact
that $\dot{E}_m = 0$ so one can include it in new parameters and use them rather than $\vc{R}_i$.
Therefore the dependence of $E_m$ on $\vc{R}$ is not required. One can calculate $E_m$ numerically
and substitute it into the formulas regarding this quantity as one more external parameter. Moreover
to find $\gamma_m$ one needs only the energy $E_m$ but not the whole spectrum as in (\ref{Bn}).
It can become an important issue if one considers partially solvable models. The requirement
$\dot{E}_m = 0$ is sufficient because otherwise one has to solve the secular equation at each
moment $t$ that is equivalent to the numerical solution of the non-stationary Schr\"{o}dinger equation
itself and therefore it makes the discussed method useless.

Let us now consider some simple applications of the proposed method. First let us see how it
works for the trivial case $n=2$. (\ref{AdiabSchroed}) reduces then to two linearly dependent
equations
\begin{eqnarray*}
(B_3 \mp B)\xi + (B_1 - iB_2) & = & 0\\
(B_1 + i B_2)\xi + (-B_3 \mp B) & = & 0
\end{eqnarray*}
Here we returned to the usual notations of the magnetic field components $B_i$ and
$\pm B = \pm |\vc{B}$ is the energy of the state $|\pm >$. Choosing one of the equations and
taking the spherical coordinates we come  to one of the relations
$$
    \xi = -\tan\vartheta/2\,\, e^{-i\varphi} \quad  \xi = \cot\vartheta/2\,\, e^{-i\varphi}.
$$
for the upper and lower sign correspondingly.  Thus these are the coordinates of stereographic
projection made from the north (south) pole of the sphere. Substituting it  into the formula for
$\omega_{\pm}(\xi )$ (see Table \ref{omega} ) and integrating over a contour $\mathcal{C}$ we get
the well known result \cite{Berry}
$$    \gamma_{\pm} = \pm\frac{1}{2}\,\, \oint\limits_{\mathcal{C}}
                 \frac{\xi^*d\xi- \xi d\xi^*}{1 + |\xi |^2} =
                 \mp\frac{1}{2}\,\, \Omega (\mathcal{C}),
$$
where $\Omega (\mathcal{C})$ is the solid angle corresponding to the closed contour $\mathcal{C}$
on the sphere.

One more example which is less trivial is a generic three-level system. The $k$-th eigenvector
$\vc{\xi}_k$ is
then two-dimensional and some trivial algebra gives for its components
\begin{equation}\label{xi312}
  \begin{array}{lcl}
  \xi_1 & = & \Delta_1/\Delta_0,  \quad \xi_2 = \Delta_2/\Delta_0,  \\
        &   &               \\
  \Delta_0 & = & (H_{11} - E_k)(H_{22} - E_k) - |H_{12}|^2 \\
        &   &               \\
  \Delta_1 & = & H_{23}H_{12} - H_{13}(H_{22} - E_k)        \\
        &   &               \\
  \Delta_2 & = & H_{13}H_{12}^* - H_{23}(H_{11} - E_k)
  \end{array}
\end{equation}
Here we have omitted where possible the index $k$. Substitution of these expressions into
(\ref{abAn2}) gives the final formula for this case. Note that the use of formula (\ref{Bn}) here would
lead to sufficient computational difficulties even after making further simplifying assumptions
\cite{Korenblit}. The proposed approach makes concrete calculations visibly easier and more compact
although the final formulas are not of esthetic value. For the illustrative purpose we take here
the case when all $H_{ij}, \,\, i\le j$ but $H_{12}$ do not depend on $t$. Then substitution of
(\ref{xi312}) into (\ref{abAn2}) gives
\begin{eqnarray}
  \omega_k(\vc{\xi}) & = & i\,C_k\,\,
   \left[\, A - D_k\,\sin (\phi_{12} + \phi_{23} - \phi_{13})\,\right]\, d\phi_{12},
                                                                 \label{omega3k} \\
                     &  &  \nonumber  \\
  C_k & = & \frac{|H_{13}||H_{23}||H_{12}|}
       {\Delta_0^2(E_k) + |\Delta_1(E_k)|^2 + |\Delta_2(E_k)|^2},  \label{C3k} \\
                     &  &   \nonumber \\
       A & = & 1/|H_{13}|^2 - 1/|H_{23}|^2,  \label{A3} \\
                     &  &   \nonumber \\
       D_k & = & H_{11} + H_{22} -2E_k      \label{D3k}
 \end{eqnarray}
where $\phi_{ij}$ are arguments of the complex numbers $H_{ij}$. As it was discussed above the condition
$\Delta_0(t) \neq 0$ is supposed to be held everywhere on $\mathcal{C}$. For other cases of the geometric phase
in the 3-level system see \cite{Tregub2}.

\subsection{Non-Abelian Wilczek--Zee phase}\label{nonab_adiab}

Now we proceed with a more general case of degenerate spectrum. Quantum computation
for this case generated by an adiabatic loop in the control manifold is determined
by the same quasi-stationary Schr\"{o}dinger equation (\ref{AdiabSchroed}) where each energy level
$E_m$ corresponds to a set of eigenstates $|m_a\!\! >,\,\, a=1,\ldots d_m$. Cyclic evolution
of the parameters results in
\begin{equation}\label{NonadEvol}
  |m_a(T)\!\! > = U_{ab}(T)\, |m_b(0)\!\!>
\end{equation}
 where the matrix$U$ is presented by a $\mathcal{P}$-ordered exponent
 \begin{equation}\label{nonAbPhase}
 \cU = \mathcal{P}\exp \left( \oint_{\mathcal{C}}\A_m\right),\qquad
 (\A_m)_{ab}=i\, <\! m_b\: |\: dm_a\! >.
 \end{equation}
In this section we generalize the proposed approach to the geometric phase computation for the
generic case of degenerate energy levels \cite{TrAd}.

 The set of  eigenvectors $\bsm{\xi}_{ma}$, $a=1,...,d_m$ must obey the equation
\begin{equation}\label{xina}
(H_{\perp}^{(d_m)} - E_m)\,\bsm{\xi}_{ma} =  h\, \bsm{c}_a.
\end{equation}
Here the matrix $H_{\perp}^{(d_m)}$ is constructed from the first $n-d_m$ lines
and columns of $H$, $\bsm{c}_a$ are arbitrary $d_m$-dimensional vectors and $h$ is
the following $(n-d_m)\times d_m$-matrix:
$$  h = - \begin{pmatrix} H_{1, n-d_m+1} & \ldots  & H_{1,n}\\
                           \vdots      & \ldots  & ...     \\
                    H_{n-d_m, n-d_m+1} & \ldots  & H_{n-d_m,n}  \end{pmatrix}
$$
Of course it has sense only if the condition
\begin{equation}\label{cond}
\det (H_{\perp}^{(d_m)}(t) - E_m) \neq 0
\end{equation}
is valid along the evolution process. The set of vectors $\bsm{\xi}_{ma}$ must
be orthogonalized by the standard Gram algorithm and after that we get the
orthonormal set of the eigenvectors $\bsm{z}_a$ (here and below we has omitted
the index $m$) in the form
\begin{equation}\label{orthset}
\bsm{z}_a = \frac{1}{\det\Gamma_{a-1}}\,
\begin{pmatrix}
                                 &              &       & \bsm{x}_1   \\
                                 & \Gamma_{a-1} &       & \vdots         \\
\scal{\bsm{\xi}_a}{\bsm{\xi}_1}  & \ldots       &
                      \scal{\bsm{\xi}_a}{\bsm{\xi}_{a-1}} & \bsm{x}_a
\end{pmatrix} ,
\end{equation}
where $\bsm{x}_b=(\bsm{\xi}_b, \bsm{c}_b)$ and $\bsm{c}_a$ is chosen to be the
standard orthogonal set \hphantom{     } \linebreak
$\bsm{c}_a=(0...\overbrace{1}^a ...0)$.
The matrices $\Gamma_a$ are determined by
\begin{equation}\label{Gamma}
\Gamma_a = \begin{pmatrix}
1+\scal{\bsm{\xi}_1}{\bsm{\xi}_1}&\ldots  & \scal{\bsm{\xi}_1}{\bsm{\xi}_a} \\
     \vdots                      & \ddots & \vdots                           \\
\scal{\bsm{\xi}_a}{\bsm{\xi}_1}  &\ldots  &1+\scal{\bsm{\xi}_a}{\bsm{\xi}_a}
\end{pmatrix} = 1 + Z_a^{\dagger}Z_a ,
\end{equation}
where the $(n-d_m)\times a$-matrix $Z_a$ consists of $a$ first lines of the
$(n-d_m)\times d_m$-matrix $Z = (H_{\perp}^{(d_m)} - E_m)^{-1}\, h$.
Using (\ref{Gamma}) and (\ref{orthset}) we come to the final expression for the
matrix-valued $1$-form \A :
\begin{equation}\label{finalA}
\A = \frac{i}{2}\;
\frac{
g_{ab}^{ij} (\bsm{\xi_j}^*d\bsm{\xi_i} - d\bsm{\xi_j}^*\bsm{\xi_i}) + 2\,\omega_{ab}
}
{  \det ( 1 + Z_{a-1}^{\dagger}Z_{a-1})\, \det ( 1 + Z_{b-1}^{\dagger}Z_{b-`})  },
\quad 1\le i \le a,\; 1\le j\le b,
\end{equation}
where
$$
 g_{ab}^{ij} =  \Gamma_a ^i\Gamma_b ^{*j}, \qquad
 \omega _{ab} = \scal{\bsm{\xi}_j}{\left. d\,\mathrm{Im}(g_{ab}^{ij})
                                                  \right|\bsm{\xi}_i}
 + \sum\limits_{i=1}^{min(a,b)}d\,\mathrm{Im}(g_{ab}^{ii}),
$$
and $\Gamma_a^i$ is the cofactor of $\bsm{\xi}_i$ in $\Gamma_a$. Note that the
change of our basis $\bsm{c}_a$ by $\bsm{c}_{a}^{\,\prime} = U_{ab}(\vlmd )\, \bsm{c}_b$
leads to a standard gauge transformation of \A\:
$$
 \A^{\prime} = U\A U^{\dagger} + i(dU)U^{\dagger}.
$$
The formula
(\ref{finalA}) is the desired expression of \A\ in terms of the matrix elements of the
Hamiltonian. It is correct if condition (\ref{cond}) is valid. It is not nevertheless
a principal restriction because $d_m$ does not depend on time due to adiabaticity
of the evolution and there is always at least one nonzero $n-d_m$-order minor of
$H$. Then, if the minor we choose vanishes somewhere on the loop $\mathcal{C}$ one
can always take local coordinates such that the techniques considered is applicable
on each segment of $\mathcal{C}$.
It should also be noted at the end of this section that the idea of physical
realization of the quantum gates based on the concrete system driven by external electromagnetic
fields  appears if one takes into account that for $A_n$
$E_{\alpha}$ can be realized by means of ordinary bosonic creation and annihilation
operators, namely $E_{\alpha}=a_i{\dagger}a_j$ for some $1\le i,j \le n$. Then
$E_{\alpha}$ represents nothing but two-mode squeezing operator. Thus the model
considered can be applied to optical HQC with $n$ laser beams (the case n=2 is
considered in \cite{Pachos1}) and the logical
gates $U_{\alpha} $ are just two-qubit transformations realized by transformation
of two laser beams.

The method presented here enables one to build in principal any computation for
HQC described by a Hamiltonian with a stationary spectrum in terms of experimentally
measured values exactly the matrix elements of the Hamiltonian. The method depends
weakly on the dimension of the qubit space which other models based on various
parameterizations of the system's evolution operator are very sensitive to.
Application of this method to a concrete physical model will be discussed
elsewhere.

\section{Non-Adiabatic Geometric  Phase}\label{nonadiab}
\subsection{Abelian Non-Adiabatic Phase}\label{ab_nonadiab}

The adiabatic condition of quantum system's evolution is strong enough to restrict sufficiently
the scope of the search for realistic candidates for practical realization of quantum computations
despite of some attractive features of the adiabatic case such as fault tolerance due to independence
of the evolution law on the details of the parameters' dynamics etc. Therefore it is desirable to
find physically relevant cases for which on the one hand this condition would be not necessary but
on the other hand the coherency in such a system would be not yet violated so that the notion of
the phase shift itself could have physical sense. As the adiabatic theorem is no longer valid
the property of universality of the system's dynamics (independence of the concrete form of the
functions $\vc{R}_i(t)$) is no more preserved and the evolution law is sufficiently more
complicated. Then one cannot hope to carry out a general approach to derivation of the corresponding
phase shift because in each case it depends on the fine details of the parameters variation. For
the same reason the phase can be called "geometric" only conditionally because geometric intuition
is no more helpful for this case e.g. the result can be represented as an integral over $t$ rather
than over a contour that expresses mathematically the thesis above.
On the other hand non-adiabatic conditional geometric phase that was theoretically predicted in
\cite{anandan} can be measured  if transitions taking place in the system do not lead to
decoherence \cite{appelt1}. Therefore it is also possible to use the corresponding
unitary operators to perform quantum calculations. This fact has been noticed
in \cite{keiji1}, \cite{keiji2} (see also \cite{keiji3}--\cite{das} for further references).

 Let us consider a parametric quantum system described by the Hamiltonian
$H({\vc{R}})$, where $\vc{R}(t)$ is a set of arbitrarily evolving parameters. We suppose
that evolution of the Hamiltonian is determined by unitary rotation (\ref{SRHevol})
Looking for particular solutions of the Schr{\"o}dinger equation (\ref{genSchroed})  (here we
supposed $\hbar =1$)
we take a rotating frame by assigning $\tilde{\psi} (t) = U(t)\,\psi(t)$ and get
in such a way
\begin{equation}\label{schr2}
  i\frac{\partial \tilde{\psi}}{\partial t} = (H_0(t) - i\,U^{\dagger}(t)\dot{U}(t) )
  \, \tilde{\psi}(t).
\end{equation}
Of course, this transformation generally does not help to solve equation
(\ref{genSchroed}) due to the fact that the algebraic structure of the coupling term
$- i\,U^{\dagger}\dot{U}$ can appear to be rather complicated and the last
generally does not commute with $H_0$. However if a receipt is known how to
evaluate the last term in (\ref{schr2}), further solution of this equation is
straightforward:
\begin{equation}\label{gensol}
   \psi (t) = e^{-i\,\phi_n(t)}\;\mathcal{T}\!\exp\left(-i\,\int_0^t\,
    U^{\dagger}(\tau)\dot{U}(\tau)\, d\tau \right)\, \psi (0) ,
\end{equation}
where $\phi_n(t) = \int_0^t\,E_n(\tau)\, d\tau$ is so called dynamic phase,
$\mathcal{T}$ denotes time-ordering and $E_n$ are elements of $H_0$ that is by
definition diagonal. Of course if there is no way to find $U^{\dagger}(\tau)\dot{U}(\tau)$,
expression (\ref{gensol}) is useless.

Let us illustrate it for the simplest case of spin $1/2$ in the non-adiabatically rotating
magnetic field \cite{appelt2}. Uniform rotation in the plane $\vartheta = const$ is represented by
\begin{equation}\label{Ht}
  H(t) = e^{\pm i\omega_R t\, \hat{J}_3}\, e^{-i\vartheta\, \hat{J}_2}\,\,(\Omega\, \hat{J}_3)
          \,\, e^{-i\vartheta\, \hat{J}_2}\, e^{\mp i\omega_R t\, \hat{J}_3}
\end{equation}
where the sign $+$ ($-$) corresponds to the left (right) polarization. Application of (\ref{schr2})
to (\ref{Ht}) gives for the Hamiltonian in the rotating frame
\begin{equation}\label{H1t}
  H_1(t) =  e^{-i\vartheta\, \hat{J}_2}\,\,(\Omega\, \hat{J}_3)
          \,\, e^{-i\vartheta\, \hat{J}_2} \pm \omega_R\, \hat{J}_3.
\end{equation}
To diagonalize Hamiltonian (\ref{H1t}) one has to apply one more rotation to it
$$
   H_2(t) = V\, H_1(t)\, V^{\dagger},\quad V = e^{i\vartheta^*\, \hat{J}_2}
$$
where the angle $\vartheta^*$ does not coincide with $\vartheta$ due to the second non-adiabatic term.
It should rather fulfill the condition
\begin{equation}\label{tetastar}
  \tan\vartheta^* = \frac{\sin\vartheta}{\cos\vartheta \pm \omega_R/\Omega}.
\end{equation}
The second term in the denominator of (\ref{tetastar}) is the measure of non-adiabaticity of
the motion. It is clear that the angle $\vartheta^*$ replaces the usual azimuthal angle $\vartheta$
in the formula for the geometric-like phase:
\begin{equation}\label{gammanonad}
  \gamma_{\pm} = \mp\, m_3\,\,2\pi\, (1 - \cos\vartheta^*)
\end{equation}
where $m_3$ is the third spin projection. Formula (\ref{gammanonad}) is a natural generalization
of the usual Berry's formula for the adiabatic case and coincides with it in the limit
$\omega_R/\Omega \rightarrow 0$. Note that the dependence of the result on $\omega_R$ reflects the
fact the phase is no longer truly geometric because $\omega_R$ characterizes the rotation velocity
and thus the velocity of the motion along the contour in the parameter space.

As an example of the application of the non-adiabatic formula above we propose a realization of quantum gates for a
concrete 4-level quantum system driven by external magnetic field \cite{TrSpect}.
Let us consider  a system of two qubits in a bosonic
environment described by the Hamiltonian
\begin{equation}\label{ham1}
 H = H_S + H_B + H_{SB},
\end{equation}
where $H_S$ is the Hamiltonian of two coupled spins
\begin{equation}\label{hamspin}
  H_S = H_S^{(0)} + H_S^{{\rm int}} = \frac{\omega_{01}}{2}\,\sigma_{z1}\otimes 1_2 +
        \frac{\omega_{02}}{2}\,1_2\otimes\sigma_{z2} +
        \frac{J}{4}\, \sigma_{z1}\otimes\sigma_{z2}  ,
\end{equation}
where $J$ is the coupling constant, $H_B$ is the Hamiltonian of the bosonic
 environment
\begin{equation}\label{hambos}
 H_B = \sum\limits_k\,\omega_{bk}(\hat{b}_k^+\hat{b}_k+1/2),
\end{equation}
and $H_{SB}$ is the Hamiltonian of the spin- enviroment interaction.
\begin{eqnarray}
  H_{SB} & = & H_{SB}^{(1)} + H_{SB}^{(2)},  \label{hamint} \\
        &        &     \nonumber \\
  H_{SB}^{(a)} & = & S_z^{(a)}\sum\limits_k\,( g_{ak}\hat{b}_k^+ +
                    g_{ak}^*\hat{b}_k) \quad a=1,2 . \label{hamsba}
\end{eqnarray}
Here
$$
   S_z^{(1)} = \sigma_{z1}\otimes 1_2,\quad  S_z^{(2)} = 1_2\otimes\sigma_{z2},
$$
$\sigma_z$ is the third Pauli matrix, $1_2$ is $2\times 2$ unit matrix,
$\hat{b}_k^+, \hat{b}_k$ are bosonic creation and annihilation operators and
$g_{ak}$ are complex constants. We assume that the two spins under consideration
are not identical so that $\omega_{01}\neq\omega_{02}$. The Hamiltonian
determined by (\ref{ham1}) -- (\ref{hamsba}) is a natural generalization of
Caldeira-Legett Hamiltonian \cite{caldeira} for the case of two non-interacting
spins.
Let such a system be placed in the magnetic field affecting the spins but not the phonon
modes. The only change to be made in the spin part (\ref{hamspin}) is the substitution
$$
     \omega_s \sigma_z \longrightarrow \vc{B}\vc{\sigma},
$$
Three components
of $\vc{B}$ represent a control set for the qubits under consideration.
Evolution of $\vc{B(t)}$ generates evolution of the reduced density matrix
$\rho_s(t)$ that describes the spin dynamics
\begin{equation}\label{rhos}
  i\,\frac{\partial \rho_s(t)}{\partial t} = H_S\,\rho_s(t),
  \quad \rho_s(t) = U(t)\rho(0)U^+(t).
\end{equation}
Thus given curve in the control space corresponds to a quantum calculation in
which each qubit is to be processed independently. To obtain such a calculation
as a function of control parameters we first recall some common issues of spin
dynamics. We consider the external magnetic field as a superposition of a
constant component and a circular polarized wave:
\begin{equation}\label{magfield}
  \vc{B} = \vc{B}_0 + \vc{B}_1 e^{i\omega_R t},
\end{equation}
where $\vc{B}_0$ is perpendicular to $\vc{B}_1$. It is well known that the
case of the circular polarization is exactly solvable. The evolution of an
individual spin corresponding to the Hamiltonian
\begin{equation}\label{spinham}
  H = - \vc{\mu}\vc{B}
\end{equation}
 is determined by (\ref{UgA3}) where $\hat{X}_{\pm},\,\, \hat{X}_3$ are replaced by
 $ S_{\pm} = S_x \pm iS_y,\,\, S_z $ correspondingly and
\begin{eqnarray}
  \zeta (t) & = & |\zeta (t)|\exp \left(i\Delta\omega t + i\alpha (t) + i\pi/2\right),
                           \label{xi1}  \\
                    &  &   \nonumber     \\
 |\zeta (t)| & = & \frac{\omega_{\bot}\sin (\Omega t/2)}{\sqrt{(\Delta\omega )^2 +
 \omega_{\bot}^2}},         \nonumber     \\
                    &  &    \nonumber     \\
\alpha (t) & = & \arctan \left( \frac{\Delta\omega}{\Omega}\tan (\Omega t/2)\right),
                             \nonumber     \\
                    &  &      \nonumber     \\
\phi (t)  & = & - \omega_{\bot}\,(\xi_1 n_2 + \xi_2 n_1), \label{phi1}
\end{eqnarray}
where $\Delta\omega = \omega_{\parallel} - \omega_R$,
$\Omega^2 = (\Delta\omega)^2 + \omega_{\bot}^2$, $\omega_{\bot}$ and
$\omega_{\parallel}$ are Rabi frequencies corresponding to $\vc{B}_0$ and $\vc{B}_1$
respectively and finally $\vc{n}$ is the unit vector along $\vc{B}_1$.

It is known \cite{appelt1} that the pure states acquire within the rotating wave
approximation a phase factor that after one complete cycle $T = 2\pi/\omega_R$ is:
\begin{equation}\label{totphase}
  |m(T)> = \exp (- i\phi_D + i\gamma )\,|m(0)>,
\end{equation}
where $m$ is the azimuthal quantum number and the phase is split in two parts:
dynamic
$$
\phi_D = 2\pi m\,\frac{\Omega}{\omega_R}\,\cos (\theta - \theta^*)
$$
and geometrical

\begin{equation}\label{nonadphase}
  \gamma = - 2\pi m\cos\theta^*,
\end{equation}
where $\cos\theta = B_0/B$ ($\vc{B} = \vc{B}_0 + \vc{B}_1$) and $\theta^*$
is determined by formula (\ref{gammanonad}).
The phase shift between the states $|\pm 1/2 >$ results then in
\begin{equation}\label{solangle}
  \Delta\phi_g = - 2\pi\cos\theta^*
\end{equation}
that is nothing but the solid angle enclosed by the closed curve
$\vc{B}(0) = \vc{B}(T)$ on the Bloch sphere. If the rotation is slow such
that $\omega_R/\Omega \rightarrow  0$ then $\theta^* \rightarrow \theta$ and
phase shift(\ref{solangle}) coincides with the usual Berry phase.

Thus the adiabaticity condition is not really necessary for obtaining of the
geometrical phase in an ensemble of spins if the decoherence
time is much greater than $T$. Therefore one can attempt to use this phase to
get quantum gates such as CNOT. Calculation of the corresponding phase factors
is rather straightforward because the free and the coupling parts of the spin
Hamiltonian commute with each other
$$
\left[H_S^{(0)}, H_S^{{\rm int}}\right] = 0.
$$
Therefore the coupling part can be diagonalized simultaneously with the free
part by applying of the transformation $U = U_1\otimes U_2$ where $U_{1,2}$ are
the diagonalizing matrices for each single-spin Hamiltonian respectively. This
simple fact together with the following obvious identity
$$
  U^{\dagger}\dot{U} = U^{\dagger}_1\dot{U}_1\otimes 1_2 +
  1_2\otimes U^{\dagger}_2\dot{U}_2
$$
the final formula for the part of the evolution operator that stands for the
non-adiabatic geometric phase
\begin{equation}\label{gate1}
  U_g = \exp (-2\pi i \cos\theta_1^*\, S_{1z})\otimes
  \exp (-2\pi i \cos\theta_2^*\, S_{2z}),
\end{equation}
where
$$
   \tan\theta_1^* = \frac{\sin\theta_1}{\cos\theta_1 + \omega_R/\Omega_1},
    \quad
    \tan\theta_2^*  = \frac{\sin\theta_2}{\cos\theta_2 + \omega_R/\Omega_2}
$$
  and
\begin{eqnarray}
   \cos\theta_1 = \omega_{01}/ \Omega_1, \quad &
   \Omega_1^2  =  \omega_{01}^2 + \omega_1^2,  \nonumber  \\
                &  \nonumber  \\
   \cos\theta_2  =  \omega_{02}/ \Omega_2, \quad  &
   \Omega_2^2  =  \omega_{02}^2 + \omega_1^2 . \nonumber
 \end{eqnarray}
Note that gate (\ref{gate1}) is symmetric with respect to the spin transposition
as it should be and does not depend on $J$ that is typical for geometrical phase
in spin systems where the phase depends only on the position drawn by the vector
\vc{B} on the Bloch sphere. As $J$ does not affect this position, it is
absent in the final result. We do not consider here the dynamic phase determining
by the factor
$$
U_d = \exp \left(-\frac{i}{\hbar}\, \hat{H}_ST\right).
$$
It is so because one can eliminate it by making use of the net effect of the
compound transformation proposed in \cite{ekert}. After this transformation
that is generated by two different specifically chosen contours the dynamic
phase acquired by the different spin states becomes the same and the geometric
phase of each state is counted twice. After that we get (up to a global phase)
the following quantum gate
\begin{equation}\label{gate2}
  U_g = \begin{pmatrix}
   e^{i(\gamma_1 + \gamma_2)}  &  0  &  0  &  0  \\
           0                  &e^{i(\gamma_1 - \gamma_2)} &  0  &  0  \\
                         0   &  0  & e^{i(-\gamma_1 + \gamma_2)} &  0  \\
                         0   &  0  &  0  & e^{-i(\gamma_1 + \gamma_2)}  \\
                           \end{pmatrix}.
\end{equation}
Thus we have constructed the quantum gate, which possess the advantage to be fault
tolerant with respect to some kinds of errors such as the error of the amplitude control
of \vc{B}. On the other hand this approach makes it possible to get rid of the
adiabaticity condition that strongly restricts the applicability of the gate.
Instead of this condition one needs some more weak one: $\tau \gg \omega_R^{-1}$,
where $\tau$ is the decoherence time.

\subsection{Non-Abelian and Non-adiabatic Phase} \label{nonab_nonadiab}

 In this section we give an example of both non-Abelian and non-adiabatic phase for a
concrete 4-level quantum system driven by external magnetic field \cite{TrNonAd}.
Let us consider a spin-$3/2$ system with quadrupole interaction.
Physically it can be thought of as a single spin-3/2 nucleus. A
coherent ensemble of such nuclei manifest geometric phase when placed in
rotating magnetic field. This phase is non-Abelian due to degenerate energy
levels with respect to the sign of the spin projection. Depending on the
experiment setup the phase can be both adiabatic as in Rb experiment by Tycko
\cite{tycko} and non-adiabatic as in Xe experiment by Appelt et al
\cite{appelt1}. This non- Abelian phase  results in mixing of $\pm 1/2$ states
in one subspace and $\pm 3/2$ in another one and thus can be regarded as a
2-qubit gate. The gate is generated by a non-Abelian effective gauge potential
${\bf A}$  that is the subject of computation in this section.

 We assume
the condition of the $^{131}Xe$ NMR experiment to be held so one does not need
 to trouble about the coherency in the system.
The last is described by
the following Hamiltonian in the frame where the magnetic field is parallel
to the z-axis ($\hbar = 1$)
\begin{equation}\label{pokham}
H_0 = \omega_0 (J_3^2 - 1/3 j(j+1)).
\end{equation}
Here and in what follows we omitted the hat symbol over all $J$'s for the sake of simplicity.
 For a spin-$3/2$ system we choose the third projection of the angular momentum in the
form
\begin{equation}
J_3 = \begin{pmatrix}
                        3/2  &  0  &  0  &  0  \\
                         0   &-3/2 &  0  &  0  \\
                         0   &  0  & 1/2 &  0  \\
                         0   &  0  &  0  &-1/2  \\  \end{pmatrix} =
  \begin{pmatrix}
   3/2 \sigma_3 & 0 \\
    0           & 1/2 \sigma_3 \\
  \end{pmatrix},
\end{equation}
Then two other projection operators are
\begin{equation}\hspace*{-24pt} J_1 =
\begin{pmatrix}
    0             &  0        &  \sqrt{3}/2  &  0           \\
    0             &  0        &       0      & \sqrt{3}/2 \\
\sqrt{3}/2        &  0        &       0      &  1           \\
     0            & \sqrt{3}/2&       1      &  0          \\
\end{pmatrix} =
  \begin{pmatrix}
     0     & \frac{\sqrt{3}}{2} \\
\frac{\sqrt{3}}{2} & \sigma_1 \\
  \end{pmatrix},
\end{equation}
\begin{equation}
\hspace*{12pt}
    J_2 = \begin{pmatrix}
        0             &  0        &       0      &  -\sqrt{3}/2  \\
        0             &  0        & \sqrt{3}/2   &       0       \\
        0             &\sqrt{3}/2 &      0       &      -i        \\
      -\sqrt{3}/2     &    0      &      i       &       0        \\
    \end{pmatrix} =
   \begin{pmatrix}
      0               & -i\frac{\sqrt{3}}{2}\, \sigma_3  \\
i\frac{\sqrt{3}}{2}\, \sigma_3  & \sigma_2               \\
   \end{pmatrix}.
\end{equation}
In the laboratory frame the Hamiltonian takes the form
\begin{equation}\label{movHam}
  H = \omega_0 ((\vc{J}\vc{n})^2 - 1/3 j(j+1)) =
       e^{-i\varphi J_3}  e^{-i\theta J_2}\, H_0 \,
       e^{i\varphi J_2}  e^{i\theta J_3}.
\end{equation}
 Rotation around the z-axis means that $\varphi =\omega_1 t$ and one should
 perform the unitary transformation

\begin{equation}\label{u1}
  |\psi > = U_1\, |\tilde\psi > ,\qquad U_1 = e^{-i\omega_1 t J_3}.
\end{equation} In the rotating frame we get
\begin{equation}\label{h1}
  H_1 = e^{-i\theta J_2}(\omega_0 J_3^2 - \omega_1\tilde J_3)e^{i\theta J_2} -
        \frac{5\omega_0}{4},
\end{equation}
where
$$
      \tilde J_3 = e^{i\theta J_2}\, J_3 \, e^{-i\theta J_2}
$$ Expression (\ref{h1}) is equivalent to

\begin{equation}\label{h1matr}
  H_1 = \begin{pmatrix}
         \omega_0 - \frac{3}{2}\omega_1 \cos\theta\,\sigma_3 &
                  \omega_1\sqrt{3}/2 \\
                  &        \\
         \omega_1\sqrt{3}/2            &  -\omega_0 - \frac{1}{2}\omega_1
                                \cos\theta\,\sigma_3  +
                                \omega_1\sin\theta\,\sigma_1\\
        \end{pmatrix}
\end{equation} It is convenient to diagonalize this matrix in two steps. First
we get rid of $\sigma_1$ in the last matrix element by performing of the
block-diagonal transformation
\begin{equation}\label{u2}
  U_2 = \diag (1, e^{-i\alpha\,\sigma_3}),
\end{equation}
where $\tan\alpha = 2\tan\theta$. Thereafter the Hamiltonian $H_1$ reads
\begin{equation}\label{h2}
 H_1 = \begin{pmatrix}
         \omega_0 - \frac{3}{2}\,\omega_1\, \cos\theta\,\sigma_3 &
                  \omega_1\,\sqrt{3}/2 \\
                        &     \\
         \omega_1\,\sqrt{3}/2            &  -\omega_0 - \frac{1}{2}\,\omega_1\,
                                \frac{\displaystyle\cos\theta}{\displaystyle\cos\alpha}\,\,
                                \sigma_3  \\
        \end{pmatrix}.
\end{equation}
At the second step we apply the transformation
\begin{equation}\label{u3}
 \begin{pmatrix}  \beta_1  & \beta_2\\
                -\beta_2^* & \beta_1^*\\
    \end{pmatrix},
\end{equation}
where $\beta_1$, $\beta_2$ are diagonal $2\times 2$ matrices that must obey the
unitarity condition
\begin{equation}\label{cnd1}
  |\beta_1|^2 + |\beta_2|^2 = 1.
\end{equation}
Supposing $\beta_{1,2}$ to be real and performing transformation (\ref{u3}) we
come to the diagonalization condition in the form
\begin{equation}\label{cnd2}
  \xi (\beta_1^2 - \beta_2^2) + (\lambda_1 - \lambda_2)\beta_1\beta_2 = 0,
\end{equation}
where $\lambda_1, \lambda_2$  are $2\times 2$ diagonal matrices and $\xi$ is
a parameter
\begin{eqnarray}\label{lmd12}
\lambda_1 &=& \omega_0 - 3/2\,\omega_1 \cos\theta\,\sigma_3, \\
\lambda_2 &=& -\omega_0 - 1/2\,\omega_1
                               \frac{\cos\theta}{\cos\alpha}\,\sigma_3\\
\xi &=& \omega_1\,\sqrt{3}/2\,\sin\theta
\end{eqnarray}
Assuming $\beta_2 = \mu\,\beta_1$ where $\mu$ is a diagonal $2\times 2$ matrix
as well we come to the following expressions for the matrix elements of $\mu$
\begin{equation}\label{mui}
  \mu_i = k_i + \sqrt{1 + k_i^2},
\end{equation}
where
$$
k_i = \frac{\Delta\lambda_i}{2\xi}, \quad
\Delta\lambda_i = \lambda_{1i} - \lambda_{2i}.
$$
Finally we get for the matrix elements of $\beta_{1,2}$
\begin{eqnarray}\label{beta12}
  \beta_{1i}^2 &=& 1/2(1 + k_i^2)^{-1/2}\left(k_i + \sqrt{1 + k_i^2}\right)^{-1}\\
  \beta_{2i}^2 &=& 1/2\left(1 + \frac{k_i}{\sqrt{1 + k_i^2}}\right)
\end{eqnarray}
Now one can evaluate the connection 1-form. It is convenient to represent it
as follows:
\begin{equation}\label{A}
 \vc{A} = i\, U^+dU = A\, d\phi = \begin{pmatrix}
            A_{3/2}          & A^{tr}\\
                       &    \\
            {\tilde A^{tr}}  & A_{1/2}  \end{pmatrix}\, d\phi,
\end{equation}
where all matrix elements of $A$ denote $2\times 2$ matrix-valued blocks,
$U = U_1U_2U_3$ and $U_i$ are determined by (\ref{u1}), (\ref{u2}),
(\ref{u3}) correspondingly. Here tilde denotes a transposed matrix.
After some algebra we get for the matrix elements of (\ref{A})
\begin{eqnarray}\label{Aij}
  A^{tr}  &=& \frac{1}{2} \beta_1\beta_2(3 - \cos\alpha)\sigma_3
   + \frac{1}{2}\sin\alpha\,\beta_2\sigma_1\beta_1 , \\
          & &    \nonumber  \\
  A_{3/2} &=& (a_{3/2} + b_{3/2}\,\sigma_3 + c_{3/2}\,\sigma_1)\, d\phi, \\
          & &    \nonumber  \\
  a_{3/2} &=& \frac{1}{4} \left(3\beta_{11}^2 - 3\beta_{12}^2 +
             \beta_{21}^2\cos\alpha - \beta_{22}^2\cos\alpha\right), \\
          & &    \nonumber  \\
  b_{3/2} &=& \frac{1}{4} \left(3\beta_{11}^2 + 3\beta_{12}^2 +
             \beta_{21}^2\cos\alpha + \beta_{22}^2\cos\alpha\right), \\
          & &    \nonumber  \\
  c_{3/2} &=& - \frac{1}{2}\,\sin\alpha\,\beta_{21}\beta_{22},\\
          & &    \nonumber  \\
  A_{1/2} &=& (a_{1/2} + b_{1/2}\,\sigma_3 + c_{1/2}\,\sigma_1)\, d\phi, \\
          & &    \nonumber  \\
  a_{1/2} &=& \frac{1}{4} \left(3\beta_{21}^2 - 3\beta_{22}^2 +
               \beta_{11}^2\cos\alpha - \beta_{12}^2\cos\alpha\right), \\
          & &    \nonumber  \\
  b_{1/2} &=& \frac{1}{4} \left(3\beta_{21}^2 + 3\beta_{22}^2 +
               \beta_{11}^2\cos\alpha + \beta_{12}^2\cos\alpha\right), \\
          & &    \nonumber  \\
  c_{1/2} &=& - \frac{1}{2}\,\sin\alpha\,\beta_{11}\beta_{12},
\end{eqnarray}
where $d\phi = \omega_1 dt$. Note that as $A$ does not depend on time, the final
solution does not require $\mathcal{T}$-ordering. It should be also emphasized
 here that in the
non-adiabatic case we discuss the term $A_{3/2}$ contains non-diagonal terms
that is not the case when the adiabaticity condition is held \cite{moody}.
Now the solution of the problem takes a particular form of (\ref{gensol}):
\begin{equation}\label{partsol}
   \psi (t) = e^{-i\,\phi_n(t)}\; e^{-i\omega_1 t\, A}\; \psi (0) ,
\end{equation}

Formula (\ref{partsol}) solves the problem of the evolution control for the system
under consideration. The resulting quantum gate is entirely determined by
$ A$ and the evolution law of the magnetic field, i.e. by a contour in the
parameter space. Of course it is always possible to choose the parameters so that
$A$ turns out to generate a 2-qubit transformation that produces a superposition
of basis states. For this reason
the gate can be thought of as a universal one \cite{Deutsch2}.
 Of course, a suitable speed of the parameters evolution
can not be reached by rotation of the sample as it took place in the experiment
by authors of \cite{appelt1}. Nevertheless it is clear that this manner of control
is not principle and one could imagine a situation where the parameters evolution
is provided by the controlling magnetic field by adding a non-stationary transverse
component. It should be also noted here that general formulas (\ref{Aij}) do not provide an
apparent way to realize CNOT gate or another common 2-qubit gate. They just give the evolution
law of the system provided that the external parameters vary as shown above. To knowledge of the authors
other examples of computation of a conditional geometric phase that would be both non-Abelian and
non-adiabatic are absent. To provide the gates of common interest
one has to invent some special case of the parameters variation which makes the generic evolution
operator more simple and transparent. This subject is out of the scope of this article.

\section{Conclusion}

The approach developed in \cite{TrAd}-- \cite{TrSpect} is to be applied in the models where it is
not possible to reduce the computation of the geometric phase to the case of 2-level system. Among
those relevant to QC one can point out e.g. the model with anisotropic Heisenberg ferromagnetism
where the exchange term in (\ref{hamspin}) is determined by a matrix of constants $J_{ab}$ rather
than by a single constant $J$. In this case the Zeeman terms $H_S^{(0)}$ no longer commute with
the exchange term $J_{ab}\, S_a\otimes S_b$ and to derive the expression for the geometric phase
 it is necessary to consider a more general case of 4-level system. The problem becomes more
 complicated also if the superfine electron-nucleus spin interaction must be taken into account.
 It is the case for the Kane model of silicon QC \cite{Kane}. The effective dimension of the system's
 Hamiltonian is then 16. It is hopeless to attempt to obtain an exact analytic expression for the
 system's dynamics which should be investigated numerically (see e.g. \cite{Wellard} ) but it is
 nevertheless possible to derive an exact formula at least for the  adiabatic phase.

 One more problem to be mentioned here is interaction with the environment. It can appear to be
 important not only for such issue as decoherence but it also can in principle contribute to the
 geometric phase. The simplest way to see it is to consider the model described by (\ref{ham1})--
 (\ref{hamsba}). If the external electromagnetic wave field can affect not only the qubits but
 the phonons as well. The phonon degrees of freedom can produce Heisenberg--Weyl-like geometric
 phase that can fill the sign of the spin projection due to electron-phonon term (\ref{hamint}),
 (\ref{hamsba}). Some more complicated interaction between the qubits and the environment makes
 it necessary to compute the geometric phase for a system with the symmetry algebra which is larger
 than $su(2)$ and cannot be reduced to the last one (in the sense of the phase derivation).

 Other field of application could be multi-beam optical schemes for quantum computations where
 several energy levels must be included in the scheme to provide two-qubit operations. Besides of
 some special cases \cite{Xiang} it can require more general methods  for the geometric phase
 computation.

\end{document}